\documentclass{llncs}%
\usepackage{amsmath}
\usepackage{amsfonts}
\usepackage{amssymb}
\usepackage{graphicx}%
\begin{document}


\title{Real-Time Data Driven Wildland Fire Modeling}
\author{Jonathan D. Beezley\inst{1,2}
\and Soham Chakraborty\inst{3}
\and Janice L. Coen\inst{2}
\and Craig C. Douglas\inst{3,5}
\and Jan Mandel\inst{1,2}
\and Anthony Vodacek\inst{4}
\and Zhen Wang\inst{4} }
\institute{University of Colorado Denver,
Denver, CO 80217-3364, USA \and National Center for
Atmospheric Research, Boulder, CO 80307-3000, USA \and University of
Kentucky, Lexington, KY 40506-0045, USA \and Rochester Institute of
Technology, Rochester, NY 14623-5603, USA \and Yale University, New
Haven, CT 06520-8285, USA}%
\maketitle


\begin{abstract}
We are developing a wildland fire model based on semi-empirical relations that
estimate the rate of spread of a surface fire
and post-frontal heat release,  coupled with WRF, the Weather Research and
Forecasting atmospheric model.
A level set method identifies the fire front.
Data are assimilated using both amplitude and position corrections using a morphing
ensemble Kalman filter.
We will use thermal images of a fire for observations that will be compared to
synthetic image based on the model state.
\end{abstract}


\section{Introduction}

We describe the current state of a new dynamic data-driven wildland fire
simulation based on real-time sensor monitoring.
The model will be driven by real-time data from airborne imagery and ground
sensors. It will be run on remote supercomputers or clusters.

An earlier summary of this project is in \cite{Mandel-2007-DDD-short}, where
we have used a~regularization approach in an ensemble Kalman filter
(EnKF) for wildfires with a fire model by reaction-diffusion-convection
partial differential equations \cite{Johns-2008-TEK,Mandel-2006-WMD}.

Level set methods were applied to empirical fire spread also in
\cite{Mallet-2007-MWF}, without a coupling with the atmosphere or data assimilation.
The Morphing EnKF \cite{Beezley-2008-MEK} is related to data assimilation by alignment
\cite{Ravela-2007-DAF}.


\section{Coupled fire-atmosphere model}
\label{sec:coupled-model}

The work reported in this section was done in \cite{Mandel-2007-DAW}, where more
details can be found.
Our code implements in a simplified numerical framework the mathematical ideas
in \cite{Clark-2004-DCA}.
Instead of tracking the fire by a custom ad hoc tracer code, it is now
represented using a level set method \cite{Osher-2003-LSM}.
Also, the fire model is now coupled with WRF, a community supported
numerical weather prediction code.
This adds capabilities to the codes used in previous work such as the
theoretical experiments of \cite{Clark-2004-DCA} and the reanalysis of a real
fire in \cite{Coen-2005-SBE}.
We use features of WRF's design to get a natural parallelization of the fire
treatment and its support for data assimilation.


\begin{figure}[t!]
\begin{center}
\hspace*{-0.7in}\includegraphics[width=6in]{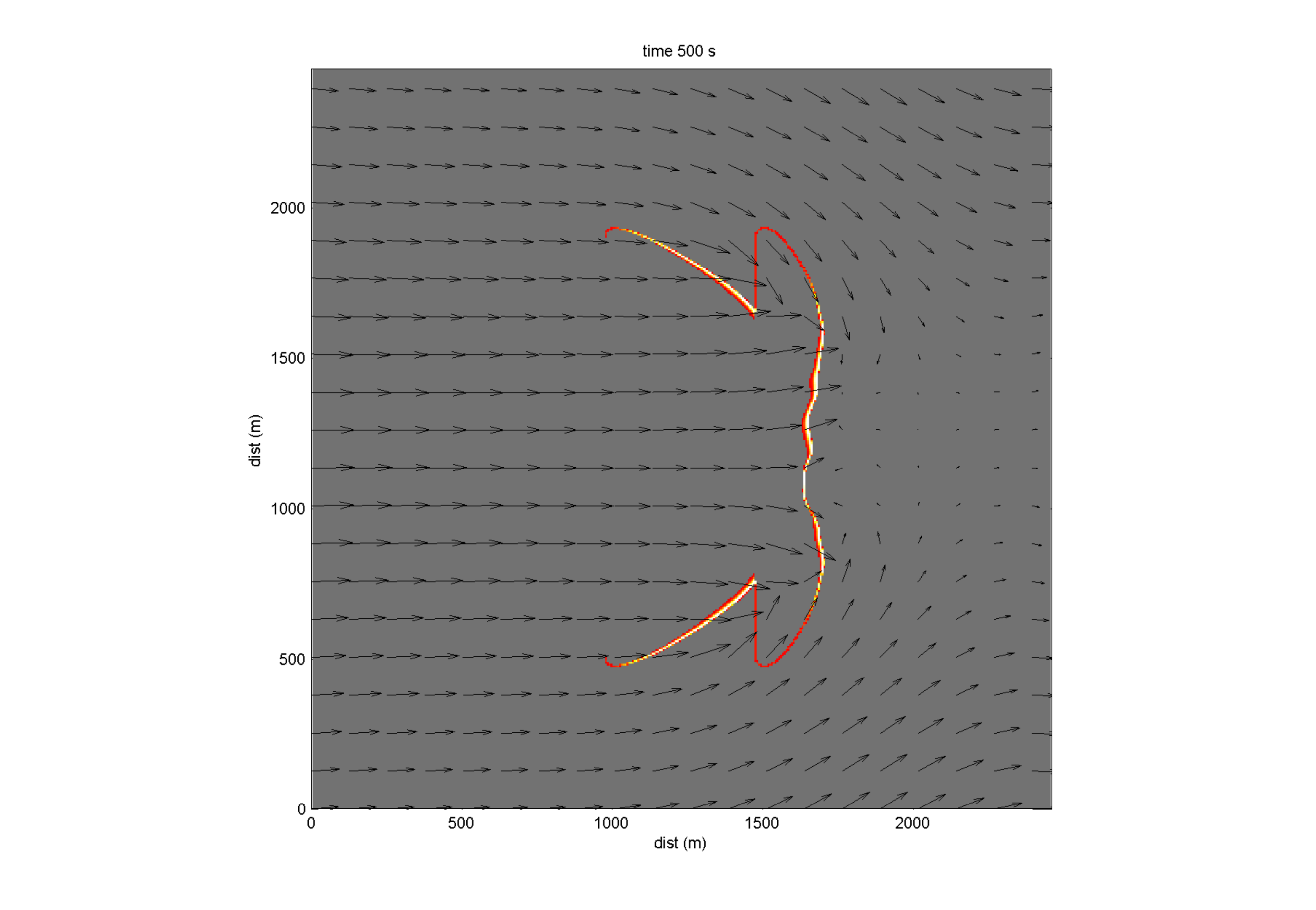}\\
\caption{Coupled fire-atmosphere simulation. Fire propagates from two line
ignitions and one circle ignition, which are in the process of merging. The
arrows are horizontal wind at ground level. False color is fire heat flux.
The fire front on the right has
irregular shape and is slowed down because of air being pulled up by the heat
created by the fire. This kind of fire behavior cannot be modeled by empirical
spread models alone.}%
\label{fig:prop}%
\end{center}
\end{figure}



We use a semi-empirical fire propagation model to represent a fire spreading
on the surface \cite{Clark-2004-DCA,Rothermel-1972-MMP} .
Given the wind speed $\overrightarrow{v}$, terrain height $z$, and
fuel coefficients  $R_{0}$, $a$,
$b,$ and $d$, determined from laboratory experiments, the
model postulates that the fireline evolves with the spread rate $S$ in the
normal direction to the fireline $\overrightarrow{n}$ given by
\begin{equation}
S=R_{0}+a\left(  \overrightarrow{v}\cdot\overrightarrow{n}\right)^b+
d\nabla z\cdot\overrightarrow{n}. \label{eq:rothermel}
\end{equation}
We further limit the spread rate to satisfy $0\leq S\leq S_{\max}$, where
$S_{\max}$ depends on the fuel.
Based on laboratory experiments, the model further assumes that after ignition
the fuel amount decreases as an exponential function of time with the
characteristic time scale depending on the fuel type, e.g., rapid mass loss in
grass and slow mass loss in larger fuel particles (e.g., downed branches).
The model delivers the sensible and the latent heat fluxes (temperature and
water vapor) from the fire to the atmosphere. The heat fluxes are taken to be
proportional to the amount of fuel burned, again with the coefficients determined
from laboratory experiments.



We use a level set method for the fire propagation.
The burning area at time $t$ is represented by the level set
\mbox{$\left\{ \left( x,y\right) :\psi\left( x,y,t\right) <0\right\}$},
where $\psi$ is called the level set function.
The fireline at time $t$ is the level set
\mbox{$\left\{ \left( x,y\right) :\psi\left( x,y,t\right) =0\right\}$}.
The level set function is initialized as the signed distance from the
fireline.

The level set function satisfies the differential equation
\cite{Osher-2003-LSM-short}
\begin{equation}\frac{\partial\psi}{\partial t}
+S\left\Vert \nabla\psi\right\Vert =0\label{eq:levelseteq}\end{equation}
which is solved numerically by the Runge-Kutta method or order two,
\begin{equation}
\psi_{n+1}=\psi_{n}+\Delta t\left[  F\left(  \psi_{n}\right)  +F\left(
\psi_{n}+\Delta tF\left(  \psi_{n}\right)  \right)  \right]  /2
\label{eq:heun}
\end{equation}
with $F\left(  \psi\right)  \approx-S\left\Vert \nabla\psi\right\Vert $,
using an approximation of $\nabla\psi$ from upwinded approximation of $\nabla\psi$
by Godunov's method \cite[p. 58]{Osher-2003-LSM-short}:
each partial derivative is approximated by the
left difference if both the left and the central differences are nonnegative,
by the right difference if both the right and the central differences are
nonpositive, and taken as zero otherwise. It is advantageous that
the propagation speed $S$ is defined by (\ref{eq:rothermel}) with
$\overrightarrow{n}=\nabla\psi/\left\Vert \nabla\psi\right\Vert$
on the whole domain, not just on the fireline, so it can be used directly
in the differential equation (\ref{eq:levelseteq}).
We use (\ref{eq:heun}) for its better conservation properties, not its accuracy.
The explicit Euler method, the obvious first choice,
systematically
overestimates $\psi$ and thus slows down fire propagation or even stops it
altogether while (\ref{eq:heun}) behaves reasonably well.



The fire model is two-way coupled with the atmospheric model,
using the horizontal wind velocity components to calculate the
fire's heat fluxes, which in turn alter atmospheric conditions such as winds.
WRF meshes can be refined only horizontally; only the finest atmospheric mesh
interfaces with the fire.
In our experiments to date, we have used time step $0.5$s with the $60$m
finest atmospheric mesh step and $6$m fire mesh step, which satisfied the CFL\
stability conditions in the fire and in the atmosphere.

Because WRF\ does not support flux boundary conditions,
the heat flux from the fire model cannot be applied directly as a boundary
condition on the derivatives of the corresponding physical field (e.g., air
temperature or water vapor contents).  Instead,
we insert the flux by modifying the temperature and water vapor concentration
over a depth of many cells with exponential decay away from the boundary.


\begin{figure*}[t!]
\begin{center}
\includegraphics[width=4.5in]{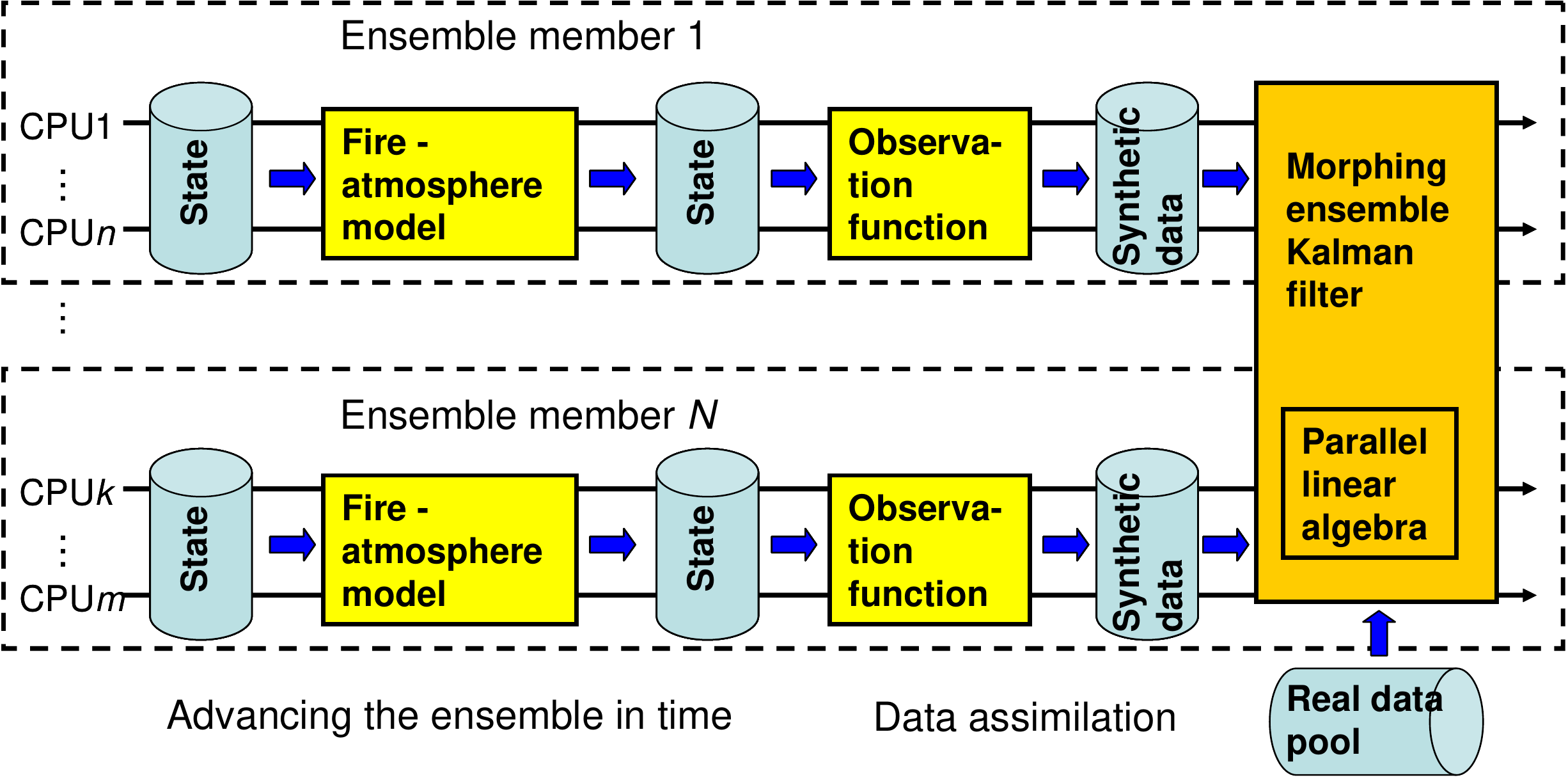}
\end{center}
\par
\caption{Parallel implementation of one assimilation cycle.
Each ensemble is advanced in time.
Using a subset of processors, the observation function is evaluated
independently for each ensemble member.
The morphing EnKF runs on all processors and adjusts the member states by
comparing the synthetic data with real data and balances the uncertainty in
the data (given as an error bound) with the uncertainty in the simulation
(computed from the spread of the whole ensemble).
Currently,
the ensemble of model states are kept in disk files.
The observation function uses the disk files and delivers synthetic
data back to disk files.
The EnKF uses the synthetic and real data and modifies the files
with the ensemble states.
The model, observation function, and EnKF are separate executables.
}%
\label{fig:parallel}%
\end{figure*}


\section{Data assimilation}
\label{sec:data-assimilation}

Our coding design is based on Fig.~\ref{fig:parallel}.
Data assimilation is done using an EnKF \cite{Evensen-2003-EKF}.
We can use data from airborne/space images and land-based sensors, though work
remains on making all the data available when needed.
In the rest of this section we describe how to treat various kinds of data,
how synthetic image data for the fire can be obtained from the system state,
and how the model states are adjusted by in response to the data.



Our efforts so far have been to get data from a variety of sources (e.g.,
weather stations, aerial images, etc.) and to compare it
\cite{Chakraborty-2008-DAV} to data from the fire-atmosphere code
\cite{Clark-2004-DCA}.
The observation function routines receives a state vector that is modified and
returned.
The state is transferred using disk files, which is slow and must be
replaced with a faster mechanism.
Individual subvectors corresponding to the most common variables are
extracted or replaced in the files.
A software layer hides both the fire code and the data
transfer method so that the code is not dependent on any particular
fire-atmosphere code.

Consider an example of a weather station that reports the location, time,
temperature, wind velocity, and humidity.
The atmosphere code has multiple nested grids.
For a given grid, we have to determine in which cell the weather station is
located, which is done using linear interpolation of the location.
The data is is determined at relevant grid points using biquadratic
interpolation.
We compare the computed results with the weather station data and then
determine if a fireline is in the cell (or neighboring ones) with the weather
station temperature to see if there really is a fire in the cell.
In the future, this will be replaced by synthetic data created from the model
state, as in Fig.~\ref{fig:parallel}.



The model state can be used to produce an image like one from an
infrared camera, which can be compared to the actual data in the data
assimilation.
This process is known as \emph{synthetic scene generation}
\cite{Wang-2008-MWF} and depends on the propagation model, parameters such as
velocity of the fire movement, as well as its output (the heat flux), and the
input (the wind speed).
We estimate the 3D flame structure using the parameters from the model, which
provides an effective geometry for simulating radiance from the fire scene.

Given the 3D flame structure, we assume we can adequately estimate the
infrared scene radiance by including three aspects of radiated energy.
These are radiation from the hot ground under the fire front and the cooling
of the ground after the fire front passes, which accounts for the heating and
cooling of the 2D surface, the direct radiation to the sensor from the 3D
flame, which accounts for the intense radiation from the flame itself, and the
radiation from the 3D flame that is reflected from the nearby ground.
This reflected radiation is most important in the near and mid-wave infrared
spectrum.
Reflected long-wave radiation is much less important because of the high
emissivity (low reflectivity) of burn scar in the long-wave infrared portion
of the spectrum \cite{Kremens-2003-MTT}.

\begin{figure}[t!]
\begin{center}
\begin{tabular}{ccc}
\includegraphics[width=1.56in]{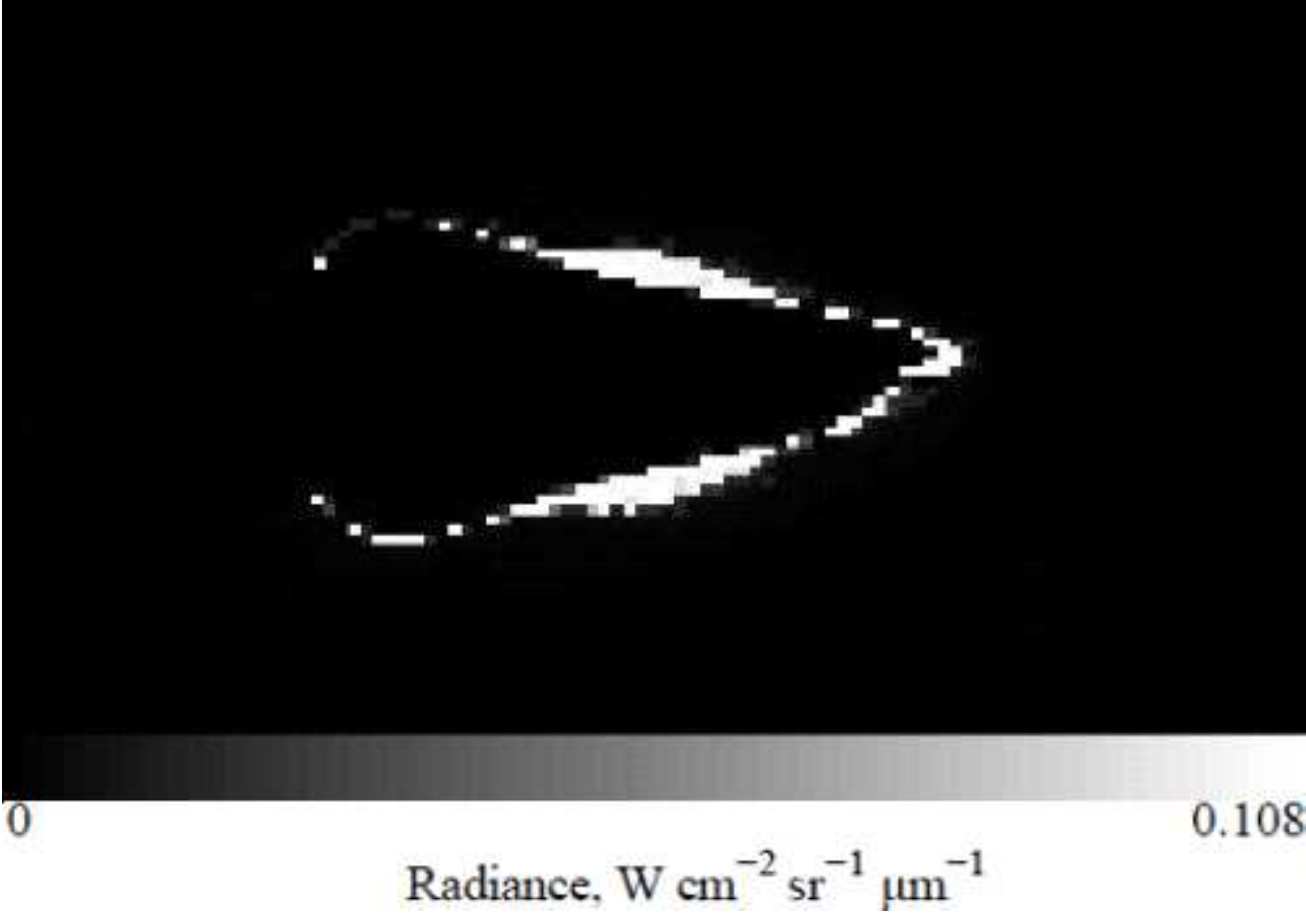} &
\includegraphics[width=1.56in]{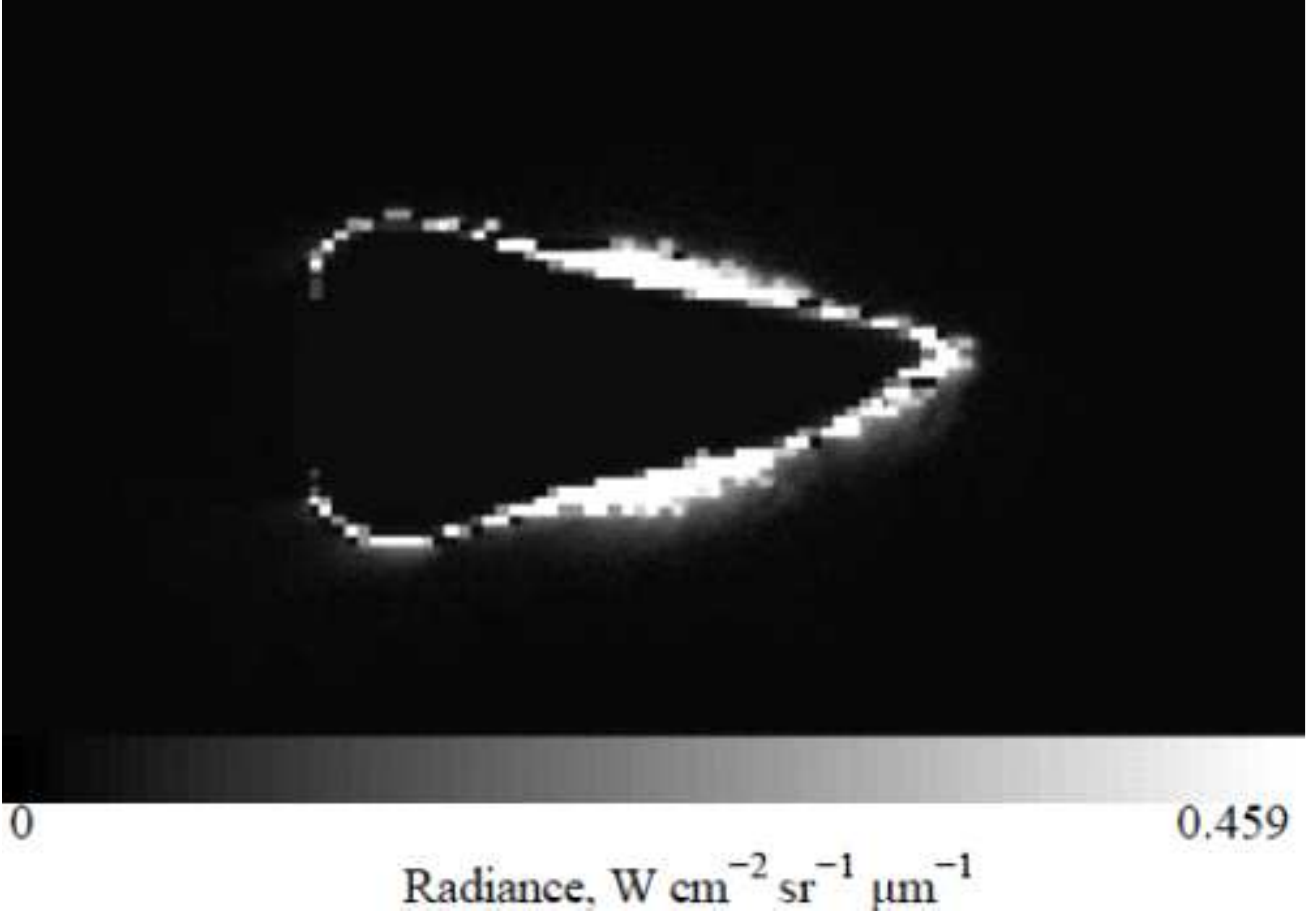} &
\includegraphics[width=1.56in]{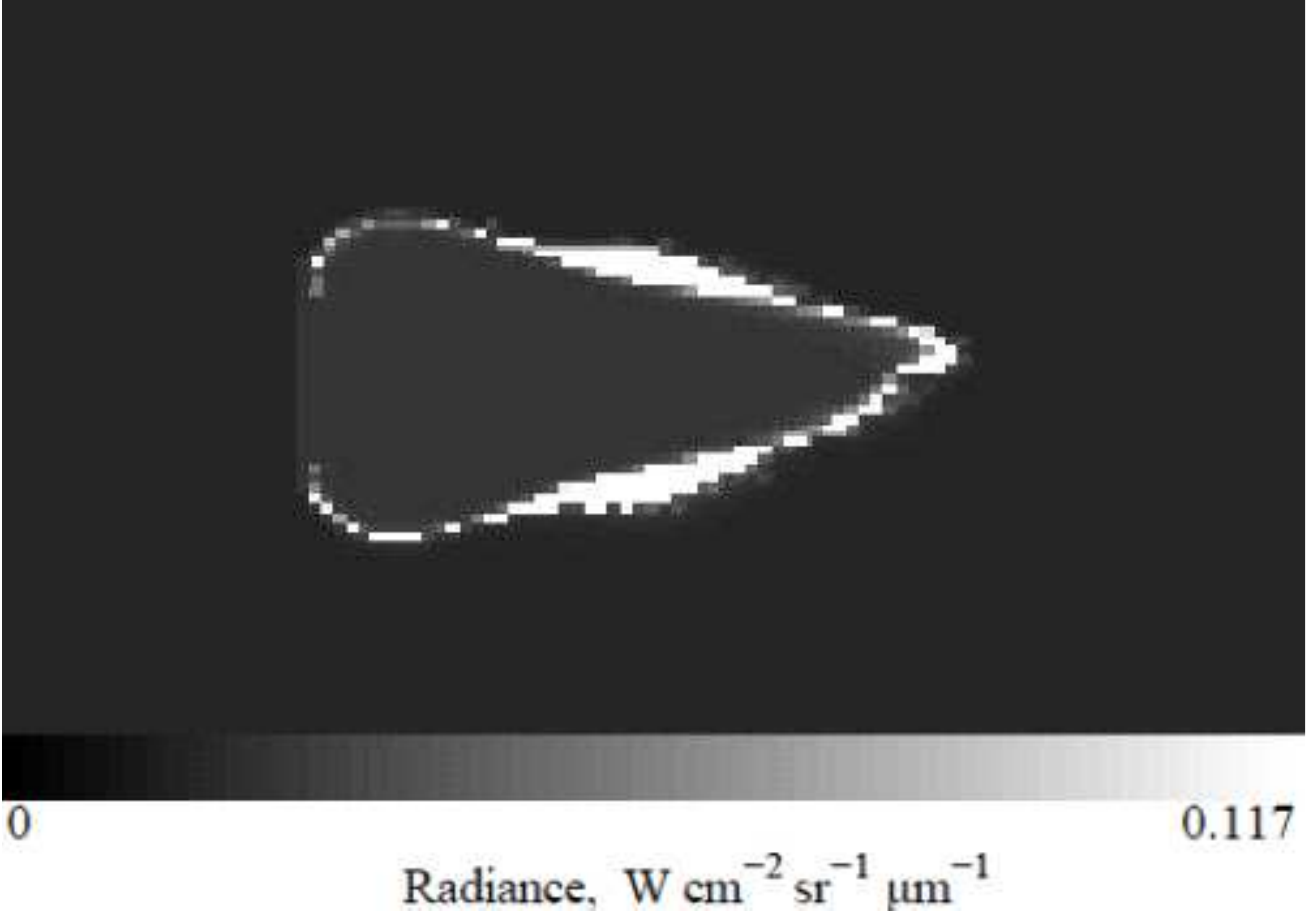} \\
(a) & (b) & (c)
\end{tabular}
\caption{Synthetic infrared nighttime radiance scenes of a simulated grassfire.
The scenes are rendered as they would be viewed by a multiwavelength camera
system on an aircraft flying 3000 m above ground.
(a) Shortwave infrared (0.9-1.8 $\mu$m).
(b) Midwave infrared (3.0-5.0 $\mu$m).
(c) Longwave infrared (8.0-11.0 $\mu$m).   From \cite{Wang-2008-MWF}.}%
\label{fig:synfire}%
\end{center}
\end{figure}

A double exponential is used to estimate the 2D fire front and cooling.
We use time constants of 75 seconds and 250 seconds and the peak temperature
at the fire front is constrained to be 1075K.
The position of the fire front is determined from the propagation model
described in the previous section.
The 3D flame structure is estimated by using the heat release rate and
experimental estimates of flame width and length and the flame is tilted based
on wind speed.
This 3D structure is represented by a 3D grid of voxels.

We use the ray tracing code Digital Imaging and Remote Sensing Image
Generation Model (DIRSIG), which is
         a first principles based synthetic image generation model developed
by the Digital Imaging and Remote Sensing Laboratory at RIT
\cite{DIRSIG-2006,Schott-1999-ASI}.
The 2D ground temperatures and the 3D voxels representing the flame are inputs
to DIRSIG, which determines the radiance from those sources as they would be
viewed by an airborne remote sensing system.
The model can produce multi- or hyper-spectral imagery from the visible
through the thermal infrared region of the electromagnetic spectrum.
The radiance calculation includes radiance reflected from the ground and the
effects of the atmosphere.
We validated the resulting synthetic radiance image (Fig.~\ref{fig:synfire})
by calculating the fire radiated energy and comparing those results to
published values derived from satellite remote sensing data of wildland
fires \cite{Wooster-2003-FRE}.
We are continuing to work on synthetic image generation with the goal of
replacing the computationally intensive and accurate ray tracing method with a
simpler method of calculating the fire radiance based upon the radiance
estimations that are inherent in the fire propagation model.



The rest of this section follows \cite{Mandel-2007-DAW} and describes a
morphing EnKF.
The model state consists of the level set function $\psi$ and the ignition
time $t_{i}$.
Both are given as arrays of values associated with grid nodes.
Unlike the tracers in \cite{Clark-2004-DCA}, our grid arrays can be modified
by data assimilation methods with ease.
Data assimilation \cite{Kalnay-AMD-2003} maintains an approximation of the
probability distribution of the state.
In each analysis cycle, the probability distribution of the state is advanced
in time and then updated from the data likelihood using Bayes theorem.
EnKF represents the probability distribution of the state by an ensemble of
states and it uses the model only as a black box without any additional coding
required.
After advancing the ensemble in time, the EnKF replaces the ensemble by its
linear combinations with the coefficients obtained by solving a least squares
problem to balance the change in the state and the difference from the data.

However, the EnKF applied directly to the model fields does not work well when
the data indicates that a fire is in a different location than what is in the
state.
This is due to the combination of such data having infinitesimally small
likelihood and the span of the ensemble does actually contain a state
consistent with the data.
Therefore, we adjust the simulation state by changing the position of the
fire, rather than just an additive correction alone, using the techniques of
registration and morphing from image processing.
Essentially, we replace the linear combinations of states in the EnKF by
intermediate states obtained by morphing, which leads to the morphing EnKF
method \cite{Beezley-2008-MEK}.

\begin{figure}[t!]
\begin{center}%
\begin{tabular}
[c]{cc}%
\hspace*{-0.1in}
\includegraphics[width=2.4in]{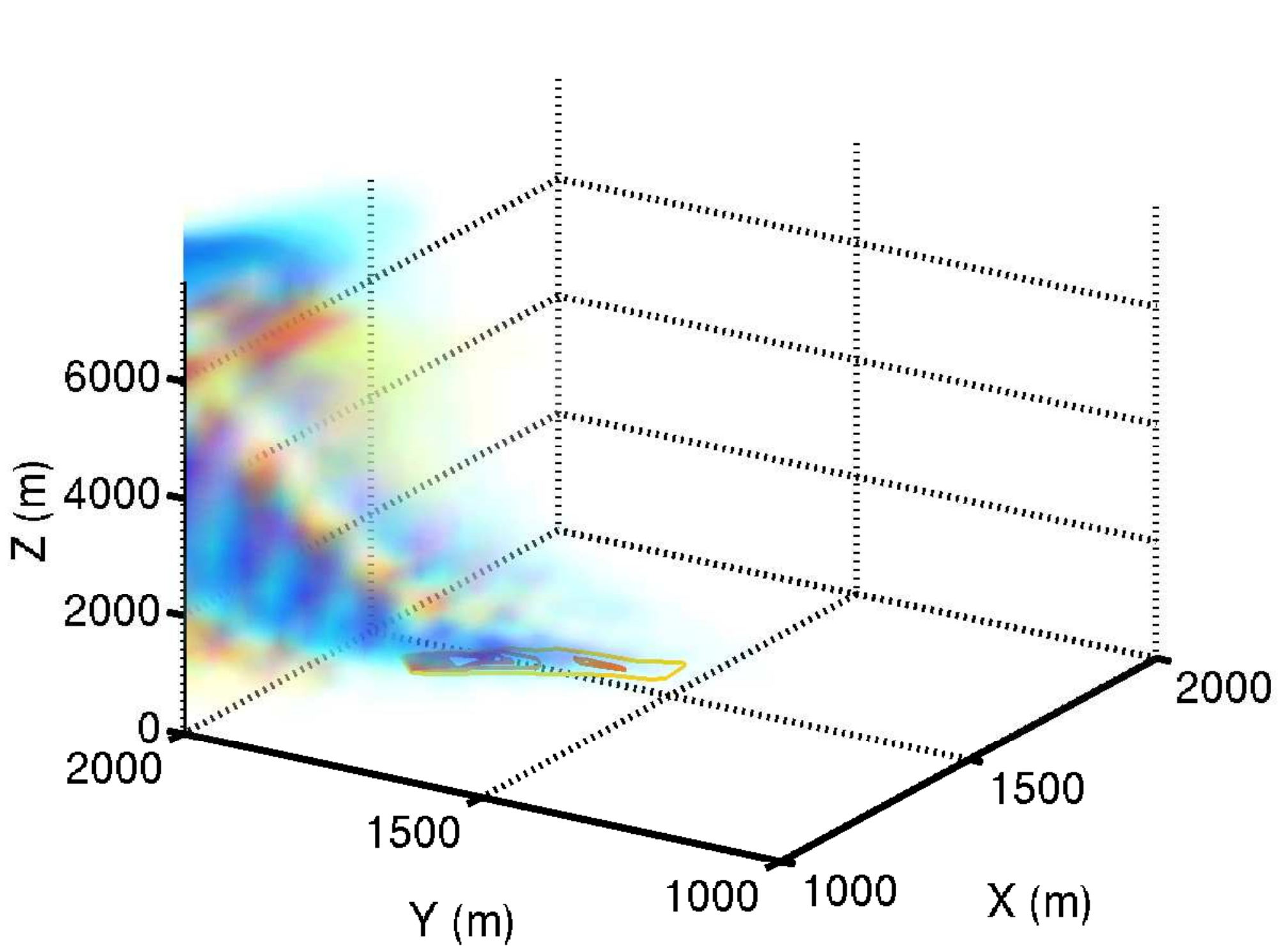}
\hspace*{-0.0in} &
\hspace*{-0.18in} \hspace*{-0.0in}
\includegraphics[width=2.4in]{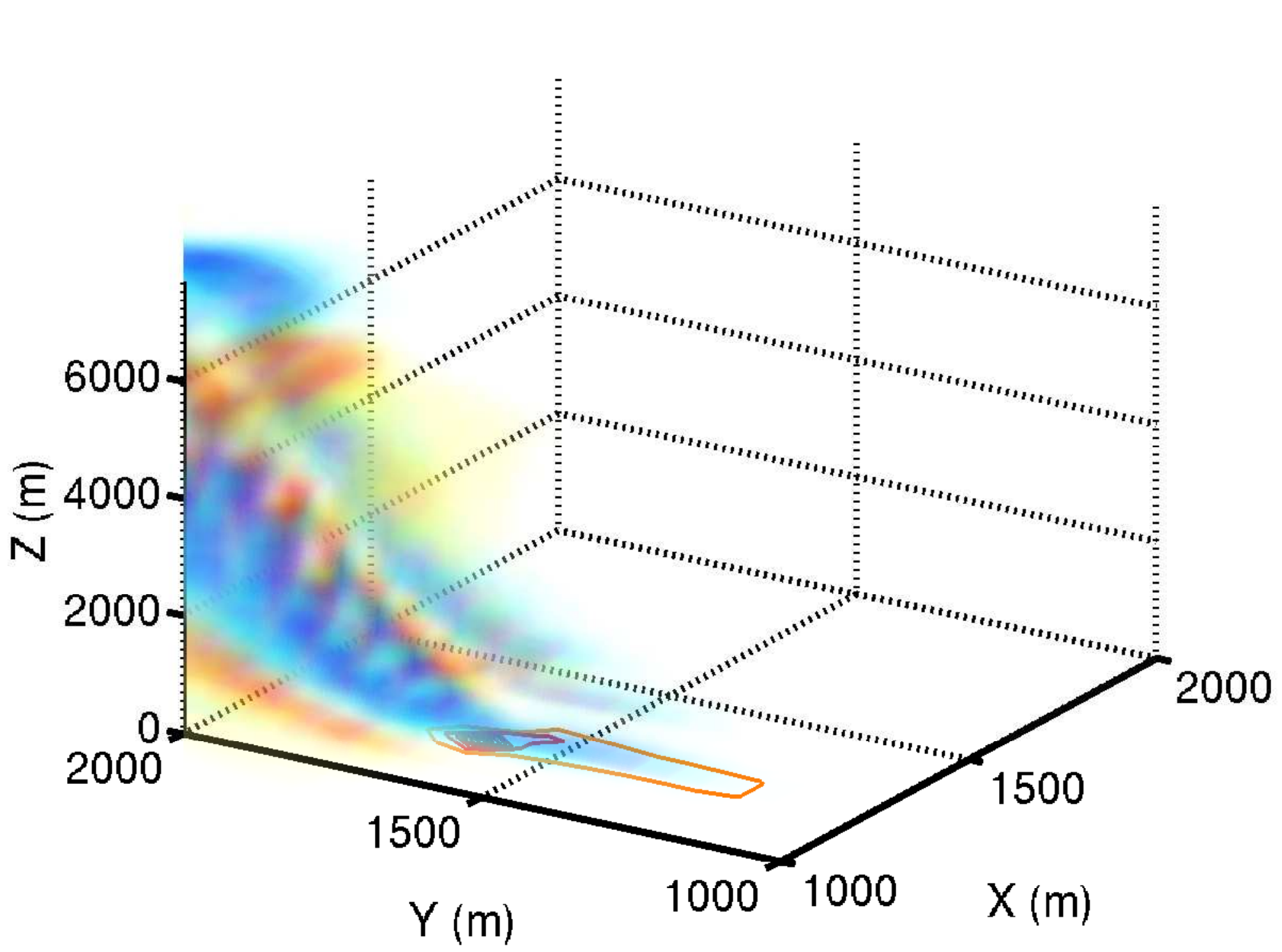}\\
(a) & (b)\\
\vspace*{-0.0in}
\hspace*{-0.1in}
\includegraphics[width=2.4in]{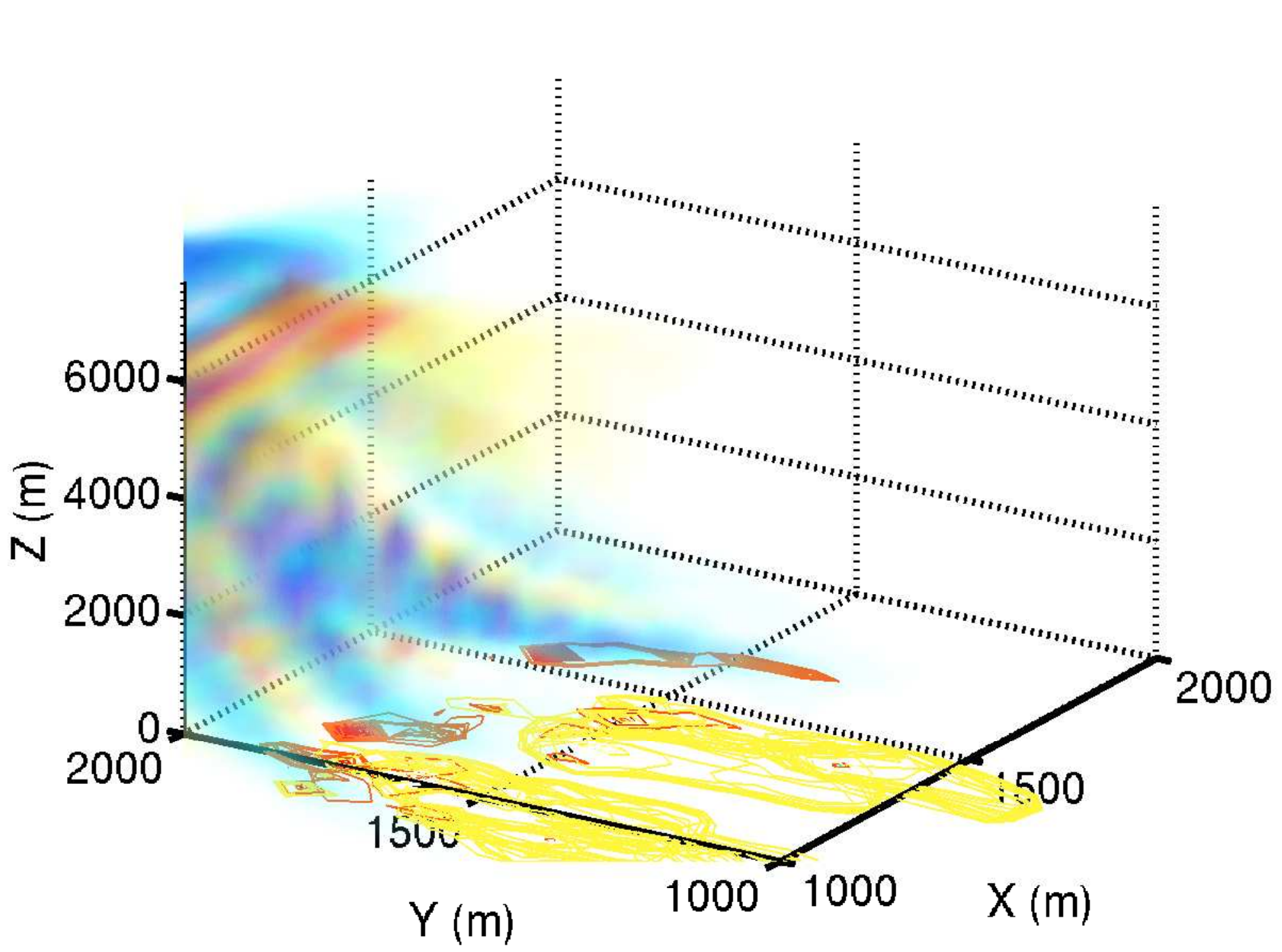}
\hspace*{-0.0in} &
\hspace*{-0.18in} \hspace*{-0.0in}
\includegraphics[width=2.4in]{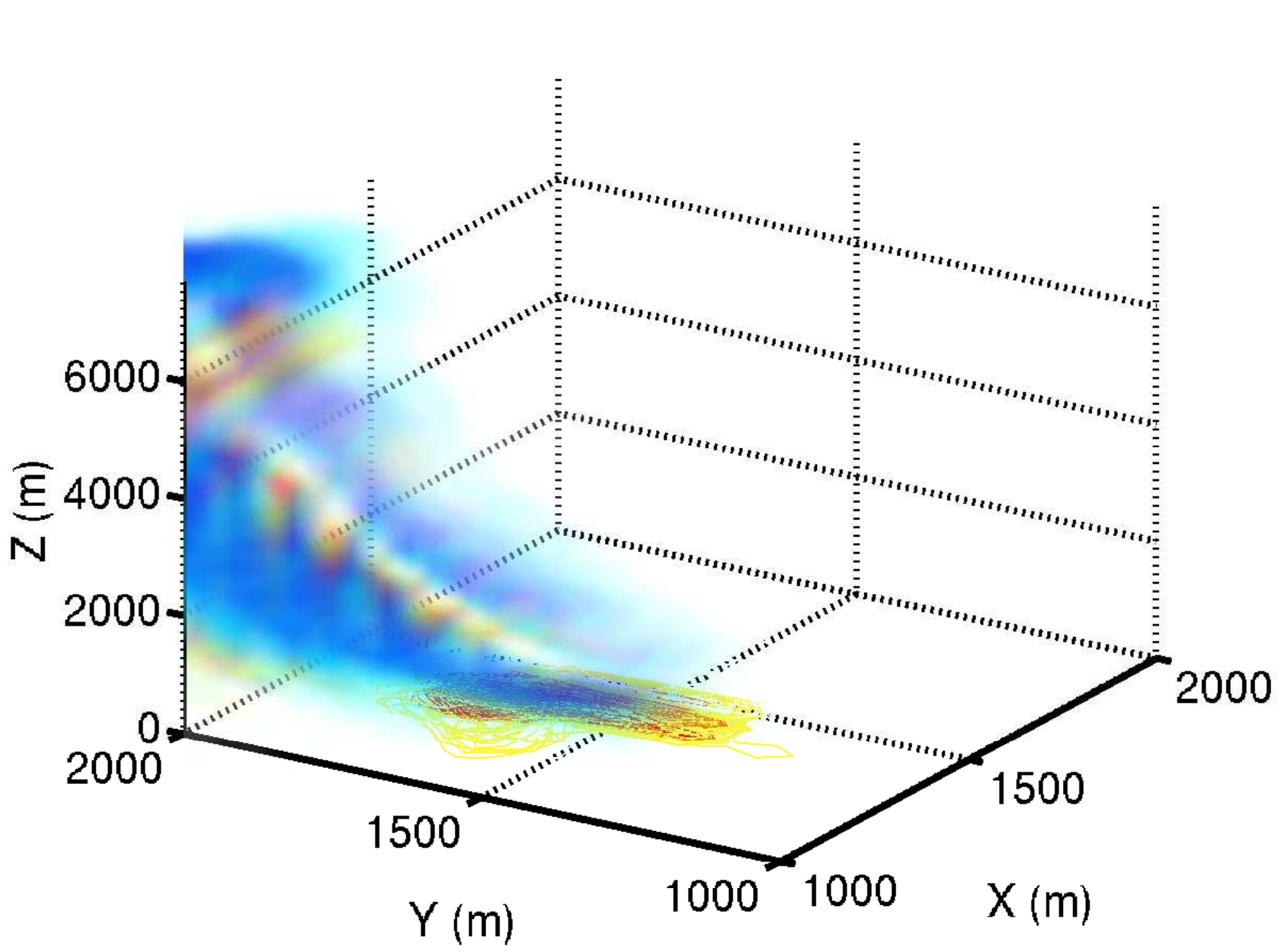}\\
\vspace*{-0.0in}
(c) & (d)
\end{tabular}
\newline
\end{center}
\caption{The morphing EnKF applied to the fireline propagation model coupled
with WRF. False color and contour on the horizontal plane is the fire heat
flux. The volume shading is the vorticity of the atmosphere. The reference
solution (a) is the simulated data. The initial ensemble was created by a
random perturbation of the comparison solution (b) with the fire ignited at
an intentionally incorrect location. The standard ENKF (c) and the morphing
EnKF (d) were applied after 15 minutes. The ensembles have 25 members each
with the heat fluxes shown superimposed. The standard EnKF ensembles diverges
from the data while the morphing EnKF ensemble keeps closer to the data.
Reproduced from \cite{Mandel-2007-DAW}.}%
\label{fig:morph-coupled}%
\end{figure}

Given two functions $u_{0}$ and $u$ representing the same physical field (e.g.,
the temperature or the level set function) from two states of the coupled
model, registration can be described as finding a mapping $T$ of the spatial
domain so that $u\approx u_{0}\circ\left(  I+T\right)  $, where $\circ$
denotes the composition of mappings and $I$ is the identity mapping.
The field $u$ and the mapping $T$ are given by their values on a grid.
To find the registration mapping \ $T$ automatically, we solve approximately
an optimization problem of the form%
\[
\left\Vert u-u_{0}\circ\left( I+T\right) \right\Vert +\left\Vert
T\right\Vert +\left\Vert \nabla T\right\Vert \rightarrow\min.
\]
We then construct intermediate functions $u_{\lambda}$ between $u_{0}$ and
$u_{1}$ using \cite{Beezley-2008-MEK}.
For $r=u\circ\left( I+T\right) ^{-1}-u_{0}$, we have
\begin{equation}
u_{\lambda}=\left( u+\lambda r\right) \circ\left( I+\lambda T\right)
,\quad0\leq\lambda\leq1. \label{eq:intermediate}%
\end{equation}
The morphing EnKF works by transforming the ensemble member into extended
states of the form $\left[ r,T\right]$, which are input into the EnKF.
The result is then converted back using (\ref{eq:intermediate}).
Fig.~\ref{fig:morph-coupled} contains an illustrative result.


\section*{Acknowledgement}

This work was supported by the NSF under grants
CNS-0325314, CNS-0324910, CNS-0324989, CNS-0324876,
CNS-0540178,
CNS-0719641, CNS-0719626, CNS-0720454,
DMS-0623983,
EIA-0219627,
and
OISE-0405349.


\bibliographystyle{splncs}
\bibliography{../../bibliography/dddas-jm}


\end{document}